\documentclass[preprint,12pt]{elsarticle}

\usepackage{amssymb}
\usepackage{amsmath}
\usepackage[utf8]{inputenc}
\usepackage[T1]{fontenc}
\usepackage{textcomp}
\usepackage{float}
\usepackage{multirow}
\journal{Journal of Computational Science}

\begin{document}

\begin{frontmatter}



\title{TwinArray Sort: An Ultrarapid Conditional Non-Comparison Based Sorting Algorithm}

\author{Amin Amini\corref{cor1}}

\cortext[cor1]{Corresponding author}
\ead{aiamini@uclan.ac.uk}

\affiliation{organization={School of Engineering and Computing, University of Central Lancashire},%Department and Organization
            addressline={Preston}, 
            city={Preston},
            postcode={PR1 2HE}, 
            state={Lancashire},
            country={UK}}

%% Abstract
\begin{abstract}
In computer science, sorting algorithms are crucial for data processing and machine learning. Large datasets and high efficiency requirements provide challenges for comparison-based algorithms like Quicksort and Merge sort, which achieve \textit{O(n log n)} time complexity. Non-comparison-based algorithms like Spreadsort and Counting Sort have memory consumption issues and a relatively high computational demand, even if they can attain linear time complexity under certain circumstances. We present TwinArray Sort, a novel conditional non-comparison-based sorting algorithm that effectively uses array indices. When it comes to worst-case time and space complexities, TwinArray Sort achieves \textit{O(n+k)}. The approach remains efficient under all settings and works well with datasets with randomly sorted, reverse-sorted, or nearly sorted distributions. TwinArray Sort can handle duplicates and optimize memory efficiently since thanks to its two auxiliary arrays for value storage and frequency counting, as well as a conditional distinct array verifier. TwinArray Sort constantly performs better than conventional algorithms, according to experimental assessments and particularly when sorting unique arrays under all data distribution scenarios. The approach is suitable for massive data processing and machine learning dataset management due to its creative use of dual auxiliary arrays and a conditional distinct array verification, which improves memory use and duplication handling. TwinArray Sort overcomes conventional sorting algorithmic constraints by combining cutting-edge methods with non-comparison-based sorting advantages. Its reliable performance in a range of data distributions makes it an adaptable and effective answer for contemporary computing requirements.
\end{abstract}

%% Graphical abstract
\begin{graphicalabstract}
\end{graphicalabstract}

%% Research highlights
\begin{highlights}
\raggedright
\item TwinArray Sort: A novel non-comparison-based sorting algorithm.
\item Efficiently uses dual auxiliary arrays for value storage and frequency counting.
\item Achieves \textit{O(n + k)} time complexity in the worst case.
\item Outperforms conventional algorithms, especially for unique arrays.
\item Suitable for massive data processing and machine learning dataset management.
\end{highlights}

%% Keywords

\end{frontmatter}

\section{Introduction and Related Work}
Sorting algorithms are essential to computer science and are used in many different applications. By rigorous optimization and study, classic comparison-based sorting algorithms like Quicksort, Merge sort, Heapsort, etc. could achieve\textit{ O(n log n)} for their average time complexities \cite{usmani2019,burnetas1997b,chaudhuri1994,hertz2005}. Nonetheless, these algorithms have inherent limitations, particularly when working with huge datasets or when technology limitations make efficiency in time and space complexity crucial. However, by using data attributes rather than direct element comparisons, non-comparison-based sorting algorithms like Bucket Sort, Radix Sort, and Counting Sort provide an alternative method of sorting arrays \cite{elkahlout2008,song2023}. As a result, these algorithms are appropriate for particular data types since they can achieve linear time complexity under certain circumstances. For example, counting the frequency of occurrence of each element in Counting Sort could result in \textit{O(n+k)} time complexity, where \textit{n} is the number of elements and \textit{k} is the range of the input values \cite{mahmoud2000}. Non-comparison-based sorting algorithms have their own set of drawbacks despite their increased efficiency. For example, Counting Sort memory usage is not ideal since it needs extra space proportionate to the range of input values, which can be a major concern for arrays with large ranges \cite{alkharabsheh2013,martinez2001}. Furthermore, there may be inefficiencies due to Counting Sort's insufficient handling of duplicate elements \cite{baber1991}. In contrast to other sorting algorithms, Bucket Sort and Radix Sort incur overhead due to their high initialization requirements and multiple passes over the data. For instance, the initialization of multiple buckets and the proper allocation of data among these buckets, would results in a substantial overhead \cite{schaffer1993,xiang2011}. The effectiveness of the algorithm in the case of Bucket Sort depends on choosing the appropriate number of buckets. When choosing too few or too many buckets, inefficiencies would result from inadequate number selection, where choosing too many, space complexity would significantly increase \cite{burnetas1997}. Furthermore, for handling data with widely changing values or huge ranges, Bucket Sort requires a large number of buckets; making it problematic and inefficient.
In order to improve efficiency and reduce time and space complexities, TwinArray Sort incorporates techniques such as dual auxiliary arrays and a conditional distinct array verifier. Its non-parametrized implementation reduces the aforementioned drawbacks significantly. TwinArray Sort is an innovative strategy that is especially beneficial for datasets with unique elements and distribution characteristics like near-sorted ordering, reversed ordering, or randomness. The TwinArray Sort is a flexible solution for a variety of sorting requirements since it can handle a broad range of data distributions with constant performance. TwinArray Sort has the potential to be used in applications like machine learning, where efficient preprocessing of large datasets could significantly impact model training times and overall performance, or data processing, where fast and efficient sorting is essential. In contrast to other algorithms, which may perform worse under specific conditions, TwinArray Sort performs well under most circumstances.

\section{Methods}
\begin{figure}[H]
\centering
\includegraphics[width=1\textwidth]{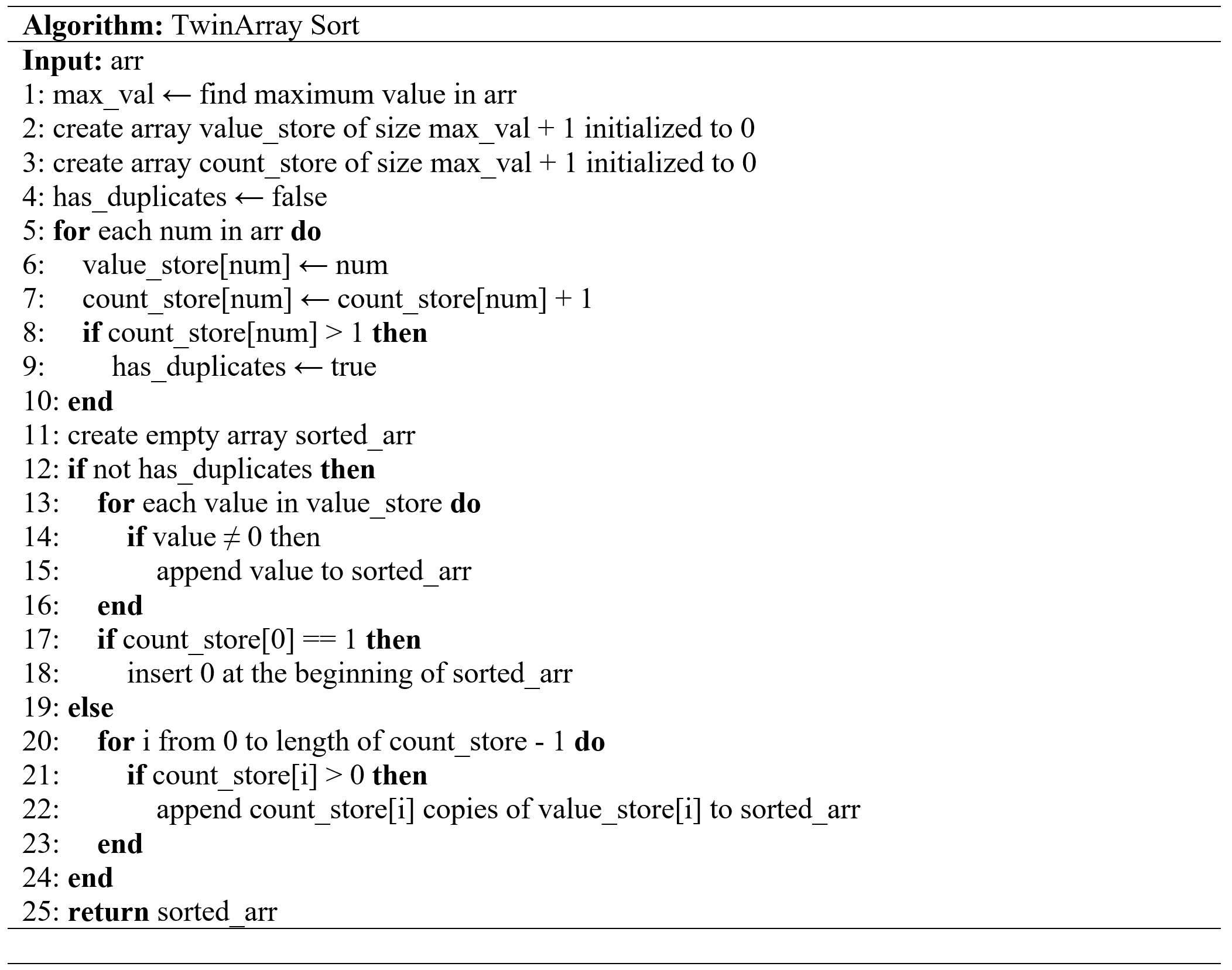}
\caption{Pseudocode of the TwinArray Sort algorithm}
\label{fig1}
\end{figure}

In order to define the size for two auxiliary arrays, TwinArray Sort first finds the maximum value within the input array. The values from the input array and the associated frequencies are kept in these arrays. Upon populating these arrays according to the indices of the arrays, the algorithm looks for duplicate elements. It creates the sorted output by either directly extracting non-zero components from the value array or by recreating elements based on their frequencies, depending on whether duplicates are discovered using its verification process. The procedure may prepend a zero if the input array’s index 0 has value 0, guaranteeing that the output array stays the same size as the input array. This technique, which works especially well with datasets that have a narrow range of integer values, successfully combines the ideas of Counting Sort with direct element insertion. Figure \ref{fig1} shows the pseudocode of the TwinArray Sort algorithm.

By utilizing the built-in indices of array elements, the TwinArray Sort method is intended to sort an array of integers in an efficient manner. This approach ensures that it can handle both unique and repeated numbers efficiently by sorting the data using dual auxiliary arrays. The TwinArray Sort algorithm is implemented and analyzed in the steps that follow.

Let $A$ be an input array with $n$ elements. $A = \{a_1, a_2, a_3, \ldots, a_n\}$, where $\forall a_i, a_j \in A, a_i \geq 0$. The TwinArray Sort algorithm involves the following steps (Figure \ref{fig2}):
\begin{figure}[H]
\centering
\includegraphics[width=1\textwidth]{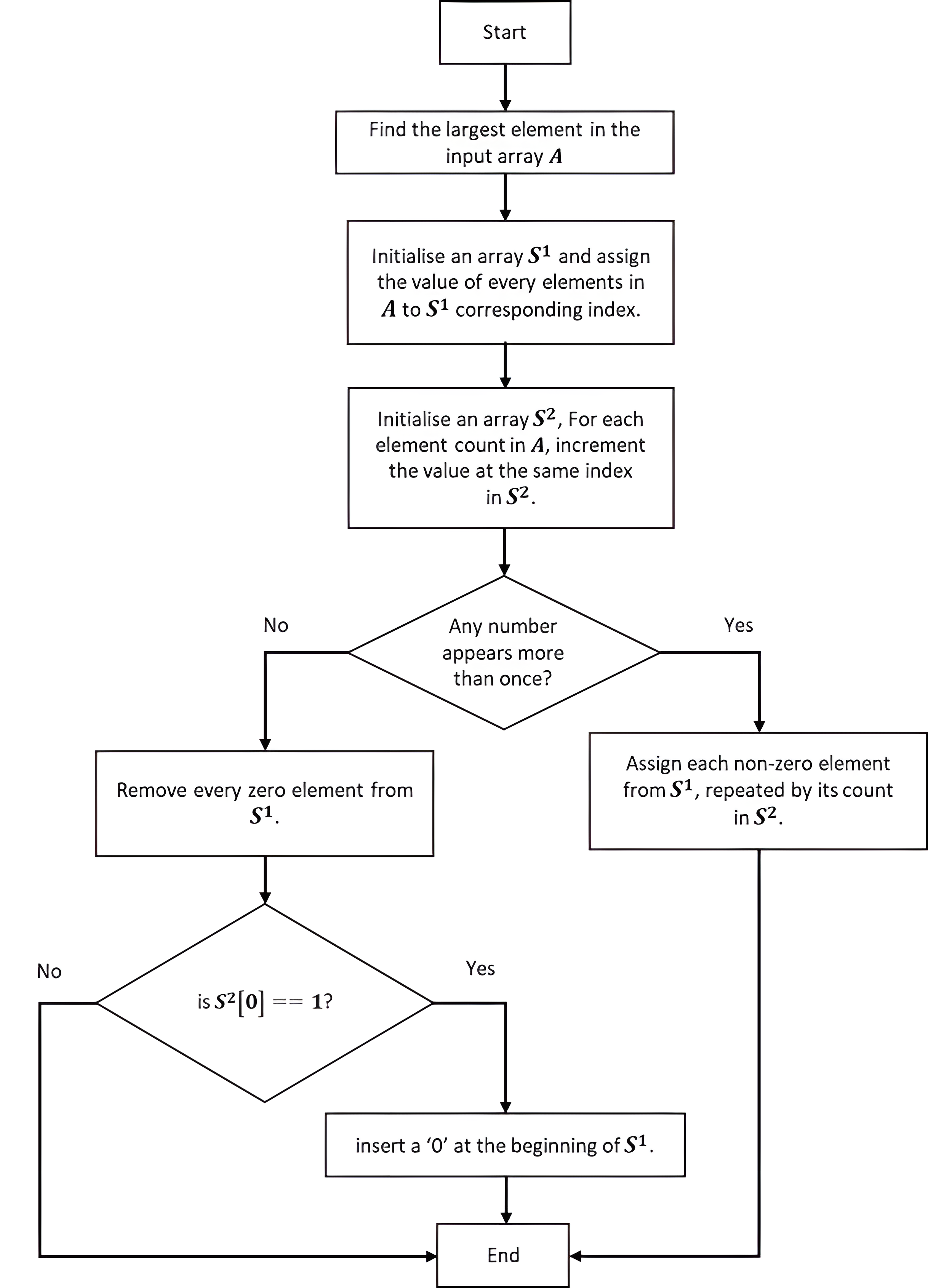}
\caption{Flowchart of the TwinArray Sort algorithm}
\label{fig2}
\end{figure}

The TwinArray Sort algorithm has an \textit{O(n+k)} time complexity, where \textit{n} is the number of elements in the input array. In order to accomplish this, one pass over the input array and one pass over the auxiliary array are made during the mapping and reconstruction phases, respectively. This would result a space complexity of the \textit{O(n+k)}, where \textit{k} is the highest value in the input array mainly due to the dual auxiliary array implementation.
While TwinArray Sort and Counting Sort have many commonalities, their methods, benefits, drawbacks, and special characteristics are very different. There are several benefits to the TwinArray Sort algorithm over Counting Sort, especially when it comes to memory optimization and effective management of duplicates. When there are no duplicates, TwinArray Sort's conditional distinct array verification mechanism helps decide whether to employ a less complex reconstruction method, which could save time in certain situations. Furthermore, TwinArray Sort divides value storage and counting using two auxiliary arrays, which may provide greater flexibility in some circumstances. TwinArray Sort has several distinctive characteristics, including independent arrays for counts and values and a conditional distinct array verifier. Unlike Counting Sort, which handles all elements equally regardless of duplicates, TwinArray Sort uses a trigger to identify whether duplicates exist and to alter sorting strategies.

\section{Experimental Setup}
For the purposes of this investigation, the TwinArray Sort algorithm was developed in Python. The effectiveness of the TwinArray Sort algorithm was assessed by comparing it to many other widely used sorting algorithms. This included Counting Sort, Pigeonhole Sort, MSD Radix Sort, Flash Sort, Tim Sort, Heap Sort, Shell Sort, Comb Sort, Bucket Sort, Block Sort, Spreadsort, Quicksort, and Merge Sort. Following a preliminary comparative investigation, the seven fastest algorithms were chosen. After that, a more thorough comparison study and analysis were conducted on these. This method made it easier to fully comprehend TwinArray Sort's performance characteristics in comparison to other widely used sorting algorithms. A variety of case studies and benchmarks were used for the comparison; these are covered in depth in the section that follows.
TwinArray Sort is contrasted in this paper with the sorting algorithms Counting Sort, Pigeonhole Sort, MSD Radix Sort, Spreadsort, Flash Sort, Bucket Sort, and Quicksort. Different data distributions based on Int64 unsigned integers were used including nearly sorted, reversed, and randomly arrays with different data sizes including \textbf{$10^5$}, \textbf{$10^6$}, \textbf{$10^7$} and \textbf{$10^8$}. The sorting algorithms' comparisons were carried out on an 8-core AMD Ryzen 7 5700X CPU with 32 GB of main memory running Ubuntu Linux on a virtual machine (WSL 2).
Selecting the middle element to serve as the QuickSort algorithm's pivot is a calculated move that strikes a balance between stability and performance. In the best-case scenario, the middle element maintains the optimal \textit{O(n log n)} time complexity by leading to roughly equal-sized partitions, which provides a decent balance \cite{roy2019}. The efficiency of the algorithm depends on this balancing keeping the recursion depth short. It is known that, in the worst-case scenario, if the center element continually splits the array incorrectly, it may result in \textit{O(n2}) time complexity \cite{lobo2020}. In spite of this, utilizing the middle element is recommended since it frequently yields a better average case than using the first or final elements, which are more likely to result in unbalanced partitions \cite{roy2019}. The middle element also tends to provide balanced partitions. Consequently, the middle element is selected in this study to help with the overall stability and performance of the QuickSort algorithm because of its capacity to reliably produce balanced partitions.
In case of Bucket Sort, it was decided to apply the number of buckets equal to the length of the array in the study’s implementation of the algorithm since this offers a balanced method for a variety of data distributions. This decision guarantees that, on average, a manageable number of items are received by each bucket, allowing for effective sorting for each bucket and keeping the overall time complexity near \textit{O(n+k)}, where \textit{k} is a small constant. When the data is equally distributed, this strategy works especially well since it creates partitions that are roughly the same size, reducing the possibility of overcrowded buckets and improving sorting efficiency \cite{xiang2011}. Additionally, handling both small and large datasets with adaptability and without the need for substantial changes or complex heuristics is made possible by using a proportional number of buckets according to the array size.
Research show that a bucket count proportionate to the dataset size can optimize sorting efficiency lends credence to the efficacy of this strategy. Research by Burnetas et al. (1997) shows that sophisticated hardware features can improve performance for small arrays sorted inside each bucket and emphasizes the significance of bucket count in establishing balanced partitions and effective sorting \cite{xiang2011}. This research matches its Bucket Sort implementation with these insights by making sure that the bucket count scales with the array size, leading to a reliable and flexible sorting solution.
Six distinct random number generators including Random, Reversed, Nsorted (nearly sorted), U\_Random (unique random), U\_Reversed (unique reversed), and U\_Nsorted (unique nearly sorted) were used to generate random arrays. Five percent of the elements in the Nsorted and U\_Nsorted arrays were displaced. To verify consistency, these identical created arrays were used to evaluate each sorting algorithm.

\subsection{Results}
A comparison of several sorting algorithms under varied input distributions and dataset sizes is shown in Table \ref{tab1}. TwinArray Sort, Counting Sort, Pigeonhole Sort, MSD Radix Sort, Spreadsort, Flashsort, Bucket Sort, and Quicksort are among the algorithms that have been examined. Arrays with distributions including Random, Reversed, Sorted, Unique Random ( U\_Random), Unique Reversed ( U\_Reversed), and Unique Sorted ( U\_Nsorted) were used to evaluate each technique. The arrays’ sizes range from \textbf{$10^5$} to \textbf{$10^8$}. The table shows the memory utilization in megabytes (MB) as well as the runtime in seconds. As it can be seen, TwinArray Sort performed significantly faster across various input distributions. For instance, TwinArray Sort completed the task in 178.6 seconds and required 2291.14 MB of memory for a dataset of size \textbf{$10^8$} with a Random distribution. In comparison to other algorithms, such as Counting Sort, which required significantly higher memory (4577.63 MB) for the same input size and distribution and ran in a much higher amount of time (487.10 s), TwinArray Sort’s performance is considerably more efficient. This is approximately 50\% of the memory requirement and is nearly 270\% faster. Similarly, TwinArray Sort was around 110\% faster and consumed 60.6\% of the memory that Pigeonhole Sort required, which ran in 196.37 seconds and uses 3814.69 MB of memory under the same conditions. It is evident from analyzing TwinArray Sort's performance in many scenarios that apart from MSD Radix Sort, it constantly used less memory compared to others. For instance, TwinArray Sort used 2332.20 MB for the Reversed distribution and dataset size of \textbf{$10^8$}, while Flashsort and Spreadsort needed 2403.25 MB and 8842.53 MB, respectively. TwinArray Sort consumes significantly less memory in this case, approximately 26.3\% of the Spreadsort memory requirements and slightly less memory than Flashsort, using about 97\% of its memory. TwinArray Sort continued to have an advantage in terms of runtime. The approach works consistently well, often surpassing competing algorithms such as Bucket Sort and Spreadsort, especially in terms of memory efficiency, even with the greatest dataset size of \textbf{$10^8$} across multiple distributions. In further comparison, TwinArray Sort performed significantly faster and more efficient than alternative sorting algorithms for unique element arrays in terms of run times and memory consumption for a wide range of input sizes and distributions thanks to its conditional distinct array verifier mechanism. For example, TwinArray Sort used 2322.32 MB of RAM to sort a uniquely distributed random (U\_Random) array of size \textbf{$10^8$} in 66.14 seconds. On the other hand, Counting Sort required much more memory (4577.63 MB) and took 510.64 seconds for the same distribution and input size. When dealing with large inputs, MSD Radix Sort performs competitively in run times (e.g., 724.03 seconds for $10^8$ size). However, its memory use is lower (1716.61 MB) than that of TwinArray Sort. While Flashsort, Bucket Sort, and Quicksort exhibit lengthier execution durations and differing memory consumption, none of them can contend with TwinArray Sort's total efficiency. When it comes to sorting unique elements across various distributions and input sizes, TwinArray Sort stands out as having the best overall mix of short run times and memory use.

\begin{table}[H]
\centering
\caption{Comparative analysis of sorting algorithms showing runtime and memory usage across different input distributions and dataset sizes}
\label{tab1}
\resizebox{\textwidth}{!}{
\begin{tabular}{|l|c|cccc|cccc|}
\hline
\textbf{Algorithms} & \textbf{Dist.} & \multicolumn{4}{c|}{\textbf{Run times (s)}} & \multicolumn{4}{c|}{\textbf{Memory (MB)}} \\
\cline{3-10}
& & \textit{n}=\textbf{$10^5$}& \textit{n}=\textbf{$10^6$}& \textit{n}=\textbf{$10^7$}& \textit{n}=\textbf{$10^8$}& \textit{n}=\textbf{$10^5$}& \textit{n}=\textbf{$10^6$}& \textit{n}=\textbf{$10^7$}& \textit{n}=\textbf{$10^8$}\\
\hline
TwinArray Sort & Random & \textbf{0.118}& \textbf{1.111}& \textbf{12.267}& \textbf{178.617}& 2.361& 23.386& 230.554& 2291.144
\\
& Reversed & \textbf{0.090}& \textbf{1.083}& \textbf{11.302}& \textbf{177.440}& 2.323& 22.923& 230.952& 2332.206
\\
& Nsorted & \textbf{0.108}& \textbf{1.104}& \textbf{11.633}& \textbf{177.939}& 2.319& 23.007& 238.275& 2347.160
\\
& U\_Random & \textbf{0.010}& \textbf{0.363}& \textbf{4.684}& \textbf{66.141}& 2.290& 23.316& 237.556& 2322.321
\\
& U\_Reversed & \textbf{0.006}& \textbf{0.069}& \textbf{0.681}& \textbf{15.337}& 2.290& 23.316& 237.556& 2322.320
\\
& U\_Nsorted & \textbf{0.011}& \textbf{0.106}& \textbf{1.213}& \textbf{19.972}& 2.290& 23.316& 237.556& 2322.320
\\
\hline
Counting Sort & Random & 0.255& 3.059& 35.237& 487.108& 4.570& 45.768& 457.757& 4577.631
\\
& Reversed & 0.237& 2.423& 25.494& 317.697& 4.572& 45.769& 457.757& 4577.631
\\
& Nsorted & 0.242& 2.521& 27.876& 369.341& 4.570& 45.769& 457.757& 4577.630
\\
& U\_Random & 0.286& 3.643& 40.947& 510.645& 4.571& 45.769& 457.757& 4577.631
\\
& U\_Reversed & 0.216& 2.167& 22.253& 244.145& 4.573& 45.769& 457.757& 4577.631
\\
& U\_Nsorted & 0.242& 2.347& 25.070& 275.164& 4.573& 45.769& 457.757& 4577.631
\\
\hline
Pigeonhole Sort & Random & 0.160& 1.631& 17.099& 196.366& 3.808& 38.140& 381.463& 3814.690
\\
& Reversed & 0.162& 1.834& 19.335& 230.885& 3.807& 38.139& 381.463& 3814.690
\\
& Nsorted & 0.160& 1.787& 19.522& 230.637& 3.808& 38.139& 381.463& 3814.691
\\
& U\_Random & 0.167& 1.980& 20.102& 239.526& 3.807& 38.140& 381.463& 3814.691
\\
& U\_Reversed & 0.171& 1.550& 15.215& 158.644& 3.807& 38.139& 381.463& 3814.691
\\
& U\_Nsorted & 0.162& 1.591& 16.169& 165.247& 3.807& 38.139& 381.463& 3814.691
\\
\hline
MSD Radix Sort & Random & 0.315& 4.181& 47.839& 663.023& \textbf{1.679}& \textbf{16.775}& \textbf{165.645}& \textbf{1650.600}\\
& Reversed & 0.240& 4.367& 49.424& 750.084& \textbf{1.653}& \textbf{16.476}& \textbf{164.225}& \textbf{1656.515}\\
& Nsorted & 0.302& 4.466& 61.134& 747.062& \textbf{1.648}& \textbf{16.636}& \textbf{165.542}& \textbf{1647.953}\\
& U\_Random & 0.323& 4.395& 58.749& 724.031& \textbf{1.718}& \textbf{17.167}& \textbf{171.664}& \textbf{1716.618}\\
& U\_Reversed & 0.319& 4.043& 53.789& 632.245& \textbf{1.717}& \textbf{17.168}& \textbf{171.664}& \textbf{1716.618}\\
& U\_Nsorted & 0.312& 4.036& 54.565& 640.853& \textbf{1.717}& \textbf{17.167}& \textbf{171.664}& \textbf{1716.618}\\
\hline
Spreadsort & Random & 0.132& 1.868& 27.347& 856.944& 8.877& 88.990& 890.985& 8842.531
\\
& Reversed & 0.124& 1.495& 26.053& 1052.850& 8.838& 88.524& 891.466& 8883.657
\\
& Nsorted & 0.188& 2.369& 27.610& 1320.442& 8.831& 88.611& 898.698& 8898.529
\\
& U\_Random & 0.224& 3.307& 35.031& 922.079& 9.917& 100.034& 1009.162& 9985.214
\\
& U\_Reversed & 0.184& 1.745& 15.993& 765.324& 9.916& 100.030& 1009.162& 9985.215
\\
& U\_Nsorted & 0.127& 1.294& 22.438& 929.743& 9.916& 100.030& 1009.166& 9985.211
\\
\hline
Flashsort & Random & 0.685& 7.443& 76.774& 910.031& 2.400& 24.031& 240.324& 2403.258
\\
& Reversed & 0.642& 7.383& 70.576& 769.748& 2.400& 24.030& 240.324& 2403.258
\\
& Nsorted & 0.662& 7.606& 73.095& 798.901& 2.400& 24.031& 240.324& 2403.258
\\
& U\_Random & 0.684& 8.158& 83.678& 935.181& 2.400& 24.031& 240.324& 2403.258
\\
& U\_Reversed & 0.646& 6.662& 67.286& 697.841& 2.400& 24.031& 240.324& 2403.258
\\
& U\_Nsorted & 0.653& 7.322& 71.148& 733.189& 2.400& 24.030& 240.324& 2403.258
\\
\hline
Bucket Sort & Random & 0.378& 5.248& 50.429& 1313.397& 8.877& 88.995& 890.985& 8842.532
\\
& Reversed & 0.347& 4.325& 48.618& 1409.581& 8.838& 88.529& 891.467& 8883.663
\\
& Nsorted & 0.436& 4.926& 49.666& 1414.722& 8.836& 88.611& 898.703& 8898.531
\\
& U\_Random & 0.487& 5.843& 61.486& 1404.084& 9.916& 100.035& 1009.167& 9985.217
\\
& U\_Reversed & 0.450& 4.393& 40.355& 1103.711& 9.916& 100.034& 1009.167& 9985.217
\\
& U\_Nsorted & 0.365& 3.640& 47.380& 1289.114& 9.916& 100.034& 1009.167& 9985.217
\\
\hline
Quicksort & Random & 0.347& 4.534& 59.453& 967.493& 3.695& 33.638& 279.969& 4682.106
\\
& Reversed & 0.271& 3.742& 52.240& 942.599& 2.435& 23.550& 241.860& 2336.865
\\
& Nsorted & 0.277& 3.839& 56.217& 946.405& 2.480& 24.284& 246.428& 2426.211
\\
& U\_Random & 0.488& 5.584& 65.363& 816.290& 5.412& 44.461& 271.089& 2598.388
\\
& U\_Reversed & 0.360& 3.320& 38.278& 472.974& 2.436& 23.550& 241.860& 2336.866
\\
& U\_Nsorted & 0.353& 3.862& 44.096& 504.749& 2.482& 24.320& 246.362& 2426.211
\\
\hline
\end{tabular}
}
\end{table}

\begin{figure}[H]
\centering
\includegraphics[width=1\textwidth]{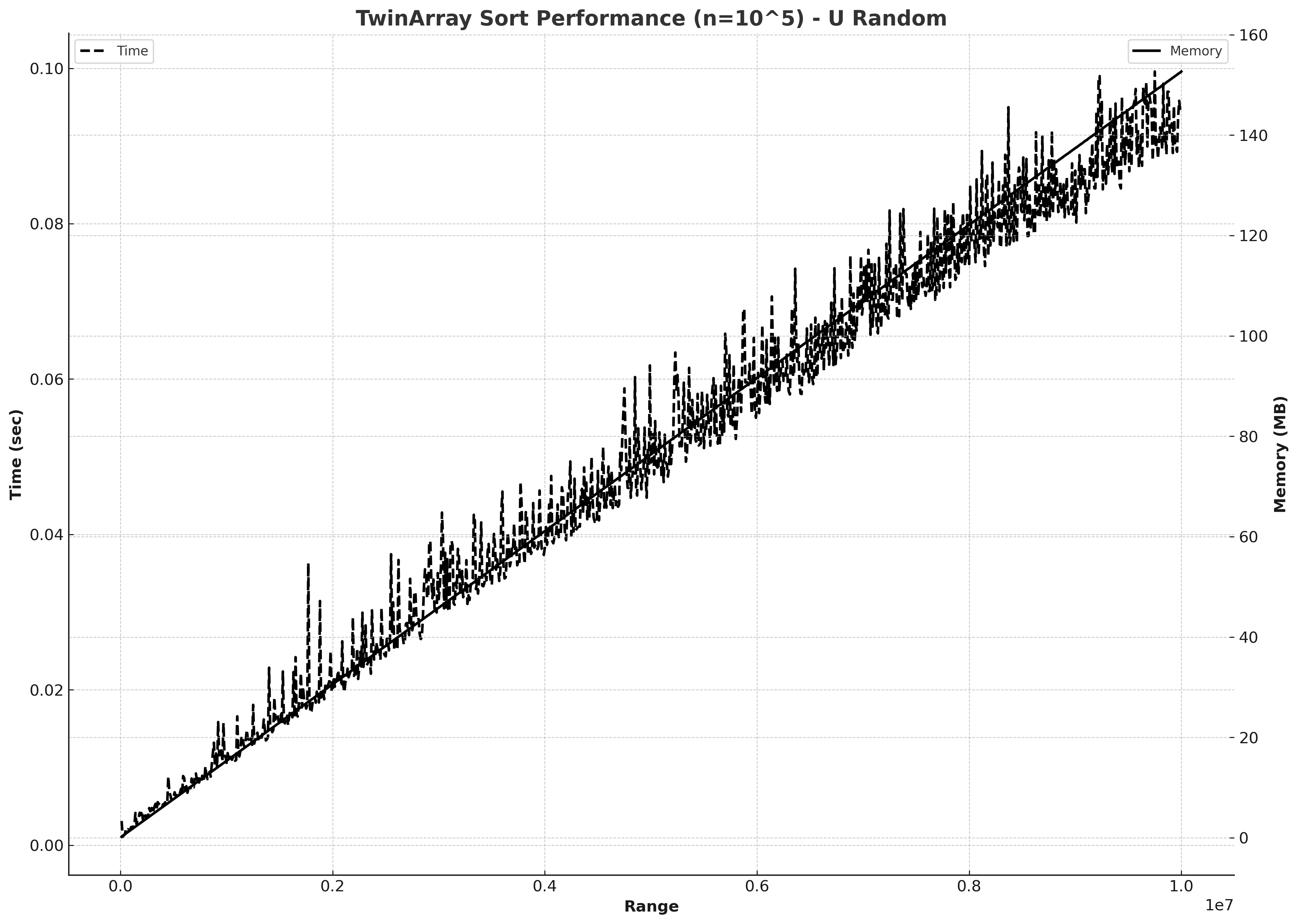}
\caption{Correlation analysis of TwinArray Sort’s time and memory usage for k >> n}
\label{fig3}
\end{figure}

It was observed that the performance of TwinArray Sort decreases, much like that of Counting sort, as the range \textit{r} grows significantly bigger than the number of elements n due to the increasing time and space complexity involved in managing a large count array. When \textit{r} is significantly larger than \textit{n}, it becomes necessary to allocate and process a big array that is mostly empty, which leads to memory waste and additional processing time. The results of the investigation (Figure \ref{fig3}) show that there is a virtually perfect positive linear relation (correlation coefficients of around 0.992 and 1.0, respectively) between the range and time and memory.

This suggests that memory use and processing time grow in a directly proportionate way as the range variable increases. In particular, the memory correlation value of 1.0 indicates that memory usage will double along with a doubling of the range. Comparably, the time taken scales roughly linearly with the range, as seen by the strong positive correlation of 0.992 for time. The graph for Time shows that there is some jitteriness in the time data. Although a perfect linear relationship, akin to memory, is what is anticipated, there are a number of reasons why this tiny deviation could occur. There may be small differences in the amount of time it takes to finish the sorting process depending on whether other processes and apps are using the computer's CPU and RAM when the sorting algorithm is running. Furthermore, Python's garbage collection feature for memory management may occasionally halt program execution in order to recover memory, leading to small discrepancies in timing measurements \cite{horsmalahti2012}. These variables add to the overall extremely linear and predictable trend, but they also cause the jitteriness in the time measurements that is shown. Thus, TwinArray Sort is less efficient in situations when the number of items is small relative to the wide range of input values because of this deterioration. It works well in situations when the range of values is proportionate to the number of elements and relatively modest.

\begin{figure}[H]
\centering
\includegraphics[width=1\textwidth]{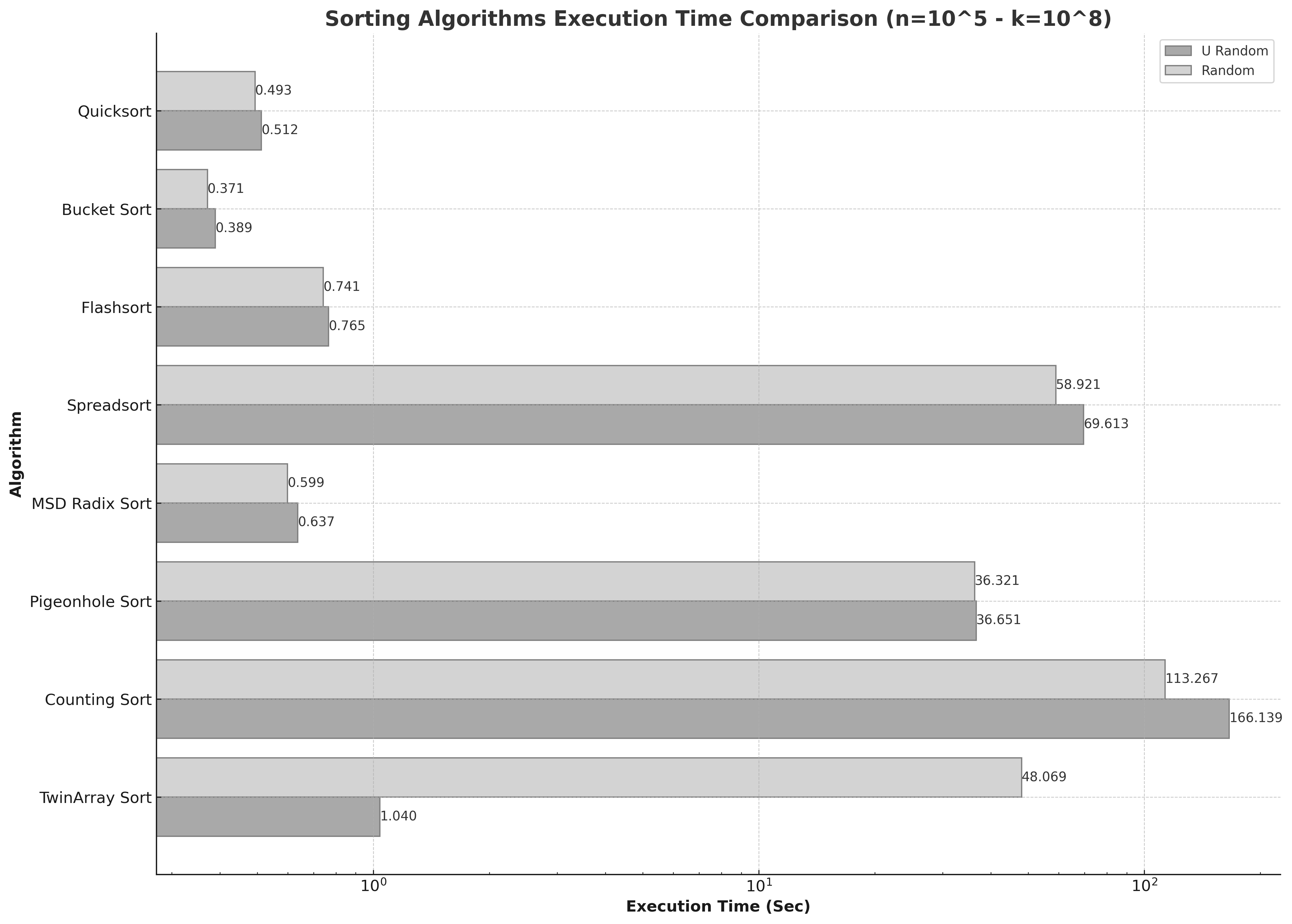}
\caption{Execution time comparison of TwinArray Sort against other sorting algorithms for both Random and U\_Random (n=\(10^5\), k=\(10^8\))}
\label{fig4}
\end{figure}

\begin{figure}[H]
\centering
\includegraphics[width=1\textwidth]{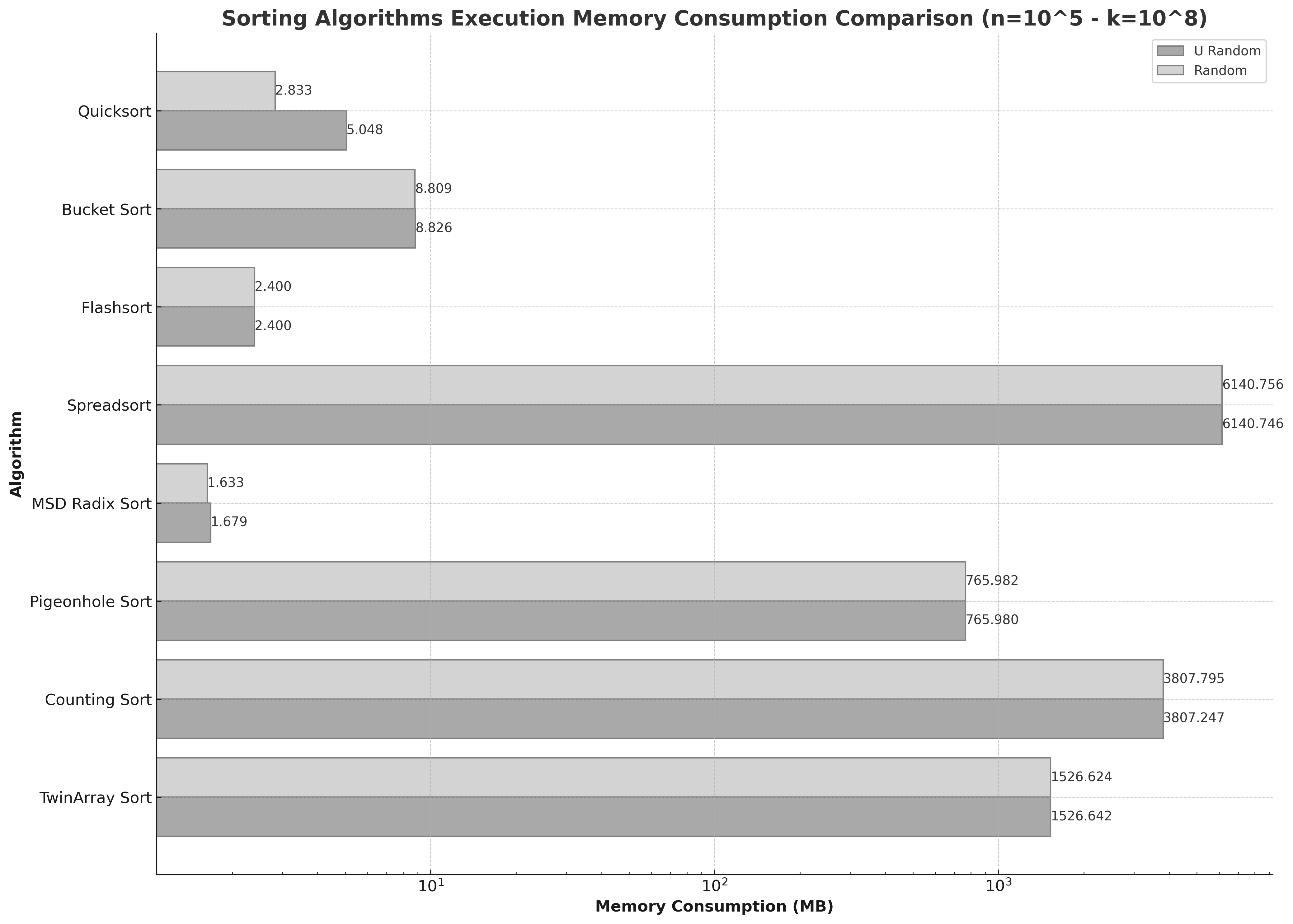}
\caption{Execution memory comparison of TwinArray Sort against other sorting algorithms for both Random and U\_Random (n=\(10^5\), k=\(10^8\))}
\label{fig5}
\end{figure}

Like other non-comparison-based sorting algorithms, the TwinArray Sort algorithm showed higher time and memory consumption when compared to other sorting techniques, as seen\textbf{ }in Figure \ref{fig4} and Figure \ref{fig5}. when the range is significantly higher than the number of elements in an array. However, TwinArray Sort performed noticeably better in terms of execution time than Spreadsort, Pigeonhole Sort, and Counting Sort when sorting datasets made up of unique numbers thanks to its conditional distinct array verifier. Furthermore, TwinArray Sort used less memory than both Spreadsort and Counting Sort, making it a more efficient method of memory consumption.

\section{Asymptotic Analysis}
Several asymptotic notations can be used to represent the temporal complexity of the TwinArray Sort algorithm in order to characterise its performance (Table \ref{tab2}). The temporal complexity, expressed in Big O notation \textit{O(n+k)}, means that the algorithm needs to process both the maximum value and every element in the array. An upper bound on the algorithm's growth rate is given by this notation. The complexity, expressed in Big Omega notation \textit{$\Omega$(n)}, indicates that in the best-case scenario, the amount of time needed is at least proportionate to the array's element count. This gives a performance bottom bound for the method. The algorithm's running time is strictly restricted by these terms, as indicated by the Big Theta notation, \textit{$\Theta$(n + k)}, which captures both the best and worst instances. TwinArray Sort is more efficient than standard comparison-based sorts for large inputs, as seen by the complexity in Little o notation of \textit{o(n log n)}, which indicates that for sufficiently large n, the growth rate of the algorithm is slower than n log n. Last but not least, the algorithm's temporal complexity in Little Omega notation \textit{$\omega$(n)}, suggests that its growth rate is quicker than any constant multiple of n and that it does not remain constant, but rather increases with the size of the input.

\begin{table}[H]
\centering
\caption{Asymptotic analysis of TwinArray Sort’s time complexity using various notations (Big O, Big Omega, Big Theta, Little o, Little omega)}
\label{tab2}
\resizebox{\textwidth}{!}{
\begin{tabular}{|l|c|l|}
\hline
\textbf{Asymptotic Notation} & \textbf{Time Complexity} & \textbf{Description} \\
\hline
Big O (O) & O(n + k) & Upper bound: The algorithm processes all elements and the maximum value. \\
\hline
Big Omega ($\Omega$)& $\Omega$(n) & Lower bound: In the best case, time is at least proportional to the number of elements. \\
\hline
Big Theta ($\Theta$)& $\Theta$(n + k) & Tight bound: Both best and worst cases involve these terms. \\
\hline
Little o (o) & o(n log n) & The algorithm grows slower than n log n for large n. \\
\hline
Little omega ($\omega$)& $\omega$(n) & The algorithm grows faster than any constant factor of n. \\
\hline
\end{tabular}
}
\end{table}

\section{Conclusion}
TwinArray Sort presents a strong substitute for conventional sorting techniques and shows notable improvements over non-comparison-based sorting algorithms. It is a very competitive technique for a wide range of datasets thanks to its effective management of duplicates, optimized memory use and conditional distinct array verification mechanism. Tests conducted on real datasets show that TwinArray Sort outperformed other sorting algorithms in all different data distributions including: randomly distributed, reversed, or nearly sorted. The algorithm's robustness and adaptability are highlighted by its worst-case \textit{O(n+k)} temporal and spatial complexities, which make it appropriate for a variety of applications. It was observed that when the range of input values k exceeds number of elements n, the algorithm's performance may deteriorate. This is because creating huge auxiliary arrays increases the complexity of both space and time. In spite of this, TwinArray Sort still outperforms Spreadsort and Counting Sort for non-unique arrays, exhibiting better memory management and duplicate handling. TwinArray Sort is the most effective option for unique arrays performs significantly faster than all other sorting algorithms, particularly due to its implementation of the conditional distinct array verification mechanism. TwinArray Sort handles unique components in a variety of distributions and input sizes with impressive scalability and performance. The technique is highly optimized for both small and large datasets, as evidenced by its continuously low run times and memory utilization. In particular, the significantly faster sorting times of reversed datasets (U\_Reversed) when compared to nearly sorted datasets (U\_Nsorted) and randomly distributed datasets (U\_Random) indicate that TwinArray Sort is very good at identifying and taking advantage of patterns in the data. Its effective memory management is demonstrated by the constant memory use across distributions and dataset sizes. TwinArray Sort offers a balanced approach between speed and memory efficiency, making it the perfect solution for applications that require the sorting of enormous unique datasets with variable distributions. In conclusion, TwinArray Sort is a flexible and effective solution for a wide range of contemporary computing applications thanks to its creative methodology and reliable performance across various data distributions and sizes. Even though TwinArray Sort’s performance may degrade when handling extremely wide input value ranges, it is still a very powerful sorting algorithm for both unique and non-unique datasets. Subsequent research endeavors may investigate refining the algorithm to alleviate its constraints and augment its relevance. TwinArray Sort is a significant advancement in sorting algorithms, fusing cutting-edge methods to overcome conventional constraints with the benefits of non-comparison-based sorting. Its reliable performance in a range of data sizes and distributions makes it an adaptable and effective answer for contemporary computing requirements.
\section*{Declaration of Interest}
The authors declare that they have no known competing financial interests or personal relationships that could have appeared to influence the work reported in this paper.

\section*{Data Availability}
Data will be made available on request.

%% References

\end{document}